\documentstyle[preprint,aps]{revtex}
\tightenlines
\begin{document}
\title{THEORY OF TRICRITICALITY FOR MISCUT SURFACES }
\author{ Somendra M. Bhattacharjee\cite{eml2}}
\address{Institute of Physics, Bhubaneswar 751 005, India}
\maketitle
\begin{abstract}
We propose a theory for the observed tricriticality in the orientational
phase diagram of Si(113) misoriented towards [001]. The systems seems to be
at or close to a very special point for long range interactions.
\end{abstract} 
\vskip 1cm

A basic question to all surface studies is the stability of a surface
against various processes, e.g., thermal fluctuations, defects, step
formation etc.  This question assumes importance especially if the surface
is cut, not in a perfect crystallographic direction, but with a slight
miscut angle (called misorientation). So far, miscut surfaces have been
found to be stable with steps or to reorganize to more complex surface
structures\cite{rev}. An exception to this general rule is the recent
observation of a {\it tricritical} point for a Si(113) surface with a small
miscut angle towards [001], point at a temperature $T_t = 1223 K$. A phase
coexistence of a stepped surface with a (113) facet, for $T<T_t$ was
observed in Ref \cite{songl,songb}.  For $T> T_t$, the surface goes
continuously to the facet as the misorientation is decreased.  In plain
English, the crystal can be cut at any small angle as one wishes so long
$T>T_t$ but, for any $T<T_t$, there is a minimum miscut angle below which
the crystal surface cannot be cut ( in thermal equilibrium).  The phase
coexistence of a reorganized facet and steps can be understood on the basis
of two separate free energy curves \cite{rev} but, a tricritial point
demands a more subtle treatment. This discovery of a tricritical point in a
two dimensional system is extremely important because it can serve as a
fertile ground for recently developed statistical mechanical
theories\cite{jjb,smbphy,lassprl,kol}.  Our aim is to describe the universal
aspects of this tricritical point.

The steps run in one direction ($``z"$ axis) without backtracking (no
overhang), so that the surface can be characterized by the density of steps,
$\rho$ (number of steps/transverse length) \cite{wortis}.  The
misorientation is related to $\rho$.  See Fig 1.  In the experiment of
Ref. \cite{songl,songb}, the coexistence curve $[\rho \sim (T_t -
T)^{\beta}$ ], was found to have a zero slope at the tricritical point, with
$\beta = 1/2$, showing utter disrespect to the extant mean field theory
\cite{songb,jaya} that predicts $\beta = 1$.

The phenomenology of the transition can be discussed in terms of the
Legendre transform of the free energy per unit transverse length, $F (\rho,
T)$ \begin{equation} {\cal F}(\mu,T; \rho) = - \mu \rho + F (\rho, T),
\label{eq:legen} \end{equation} where $\mu$ is the chemical potential for
steps.  The thermodynamic value of $\rho$ comes from $\partial {\cal
F}/\partial\rho = 0$, or equivalently $\partial F/\partial \rho = \mu$.
Taking $f_0$ as the free energy of an isolated step, Eq. (1) can be
expressed as \begin{equation} {\cal F}(\mu,T; \rho) = (f_0 - \mu) \rho +
F_{int} (\rho, T), \label{eq:free} \end{equation} where $F_{int} (\rho, T)$
is the free energy contribution from interactions among the steps.  For
noninteracting steps, $F_{int} = 0$.  Therefore, a first order transition at
$\mu=f_0$ takes a facet $(\rho=0)$ to a fully stepped surface for $\mu >
f_0$.  For purely repulsive (``fermionic") steps, $F_{int} \sim \rho^3$
leading to the continuous Pokrovsky-Talapov (or ``3/2'' order)
\cite{pt,nag75,wortis,dennijs} transition with $\rho \sim \mid f_0 - \mu
\mid^{1/2}$.  This has been the rule for almost all systems until the
tricriticality in Si(113) was discovered.

The occurrence of a phase separation, as noted in ref \cite{songb}, suggests
the existence of attractive interactions among the steps.  The steps are
generally taken to be nonintersecting.  In addition, they are expected to
have dipolar or elastic long range $r^{-2}$ interaction.  A long range
$(r^{-2})$ attraction for the non-intersecting steps, in a mean field
(``Hartree Fock") analysis, gives $F_{int} \sim $ $(T - T_t) \rho^3 + a
\rho^4$, predicting a tricritical point with a linear phase boundary
\cite{jaya}.  With an attractive $r^{-2}$ interaction, this mean field form
of $F_{int}$ agrees, in the domain of overlap of parameters, with the exact
Bethe ansatz solution \cite{suth}.  This rules out the possibility of
fluctuations causing a zero slope phase boundary in this model with long
range attration.

We therefore consider a completely different scenario. Our proposal is that
the tricriticality occurs when the steps have a short range attraction. The
physical picture we have in mind has strong resemblance to a phase
separating polymer solution.  In fact taking the steps as directed polymers
(DP), the stepped face can be thought of as a DP solution.  Our proposal is
that in the high temperature phase, the steps are repulsive and the phase is
dominated by the entropic interaction.  As the tricritical point is
approached, the attractive part starts playing a role.  The steps start
colliding and the average separation between collisions determine the
correlation length.  When this length becomes comparable to the separation
of the steps the tricritical domain is reached. The phase separation takes
place in this regime before any bound state can form.

To study the phase separation and the coexistence curve, we use a canonical
ensemble approach, and use the analogy with polymer solution \cite{dupla}.
The phase boundary is identified by equating the ``osmotic" pressure of the
two coexisting phases.  The``osmotic" pressure in this context, would mean
the excess pressure generated by the addition of one more step and is
obtained from $\Pi = \rho^2 \partial/\partial \rho(F/\rho)$, where $F(\rho)$
is the free energy as a function of density $\rho$.  Since one phase is a
flat surface with zero density, its osmotic pressure is zero.  The
coexistence curve is therefore obtained from \begin{equation} \partial \left
( \rho^{-1}F\right )/{\partial \rho} = 0.\label{eq:osmo} \end{equation}

It is the interaction that determines the shape of the coexistence curve.
We, therefore, consider two different possibilities: tricriticality with (i)
short range interactions, and (ii) long range interactions.  It seems that
the latter holds the key.

The general approach is to start from a mean field or effective free energy.
Renormalization group (RG) approach is then used to incorporate the effects
of fluctuations.  The RG $\beta$ functions tell us the effective couplings
as the length scales are changed.  Integrating the RG equations, one can
then obtain the renormalized interactions or coupling constants for the
relevant length scale $\sim \rho^{-1}$.  These renormalized coupling
constants can then be used in the mean field free energy to get $F$ for
Eq. \ref{eq:osmo}. This is justified because we are interested not in the
details of the tricritical behavior but rather in the phase boundary where
all length scales remain finite.

To write the Hamiltonian, we note that short range attractions in DP's lead
to bound states for two isolated steps.  So far as the binding transition is
concerned in low dimensions, the universal critical behaviour is independent
of any further details like noncrossing condition of the steps.  This is
known from exact renormalization group analysis and simple quantum
mechanical calculations\cite{jjphysa,sutphys,lipow}. Treating the steps as
structureless wandering lines, the Hamiltonian in a continuum formulation is
taken as \begin{mathletters} \begin{equation} H= \int dz \left [ \frac{1}{2}
\sum_i \left ( \frac{\partial {\bf r}_i}{\partial z} \right )^{^2} + v_2 \
\sum_{i>j} \delta({\bf r}_{ij}(z)) \right ] + H_{int},\label{eq:hama}
\end{equation} where ${\bf r}_i(z)$ is the $d$ dimensional transverse
position of the $i$th step at a coordinate $z$ measured along the step from
one end, ${\bf r}_{ij}(z) = {\bf r}_j(z) - {\bf r}_i(z)$ is the separation
between two steps $i$ and $j$, and $v_2 = v_{20}(T-T_t)$ is the effective
two step contact (short range) interaction. $H_{int}$ is the additional
interaction, and two possible choices are \begin{equation} H_{int} =
 v_3 \int \sum
\delta({\bf r}_{ij}(z)) \delta({\bf r}_{ik}(z)) dz,\label{eq:hamc}
\end{equation}
or,
\begin{equation}
H_{int} = h \sum_{i<j} \int \ \mid {\bf r}_{ij}(z)\mid^{-2}
dz.\label{eq:hamb} 
\end{equation}
\end{mathletters}
The first form represents a three step contact
repulsion  while the last form represents a two step long range repulsive
interaction.   For $T>T_t$, $v_2 >0$ and the noncrossing condition is
ensured by taking the limit $v_2 \rightarrow \infty$.  This however is
not required because it is known from RG that the repulsive case is 
described for $d<2$ by a stable fixed point (FP) (see below).

In a mean field treatment $\sum \delta(r_{ij})$ in Eq.  \ref{eq:hama} can be
replaced by $\rho^2$.  Similarly the three body interaction would generate a
$\rho^3$ term \cite{jjb,smbphy,lassprl}, as also the $r^{-2}$ repulsive
interaction \cite{jaya,suth} so that the mean field free energy is
\begin{equation} {\cal F}_{int} (\rho,T) = f_0 \rho + v_{20}\ (T- T_t)
\rho^2 + c \rho^3, \label{eq:mffree} \end{equation} where $c$ depends on $h$
or $v_3$ as the case may be.  This again gives $\beta = 1$, when
Eq. \ref{eq:osmo} is used, though the physics behind this is completely
different from that proposed is Ref. \cite{songl,songb,jaya}.

Let us first consider the short range case, Eq. \ref{eq:hamc}.  A simple
dimensional analysis shows that $v_3$ is marginal in $d=1$.  We introduce
the dimensionless parameters $u_2 = v_2 L^{2-d}$ and $u_3 = v_3 L^{1-d}$
where $L$ is an arbitrary length scale in the transverse direction. A
renormalization procedure would take into account the effects of
interactions at scales $< L$ along the steps, changing the effective
interaction felt at length scale $L$ (``coarse graining''). The details can
be found in Refs.  \cite{jjphysa,jjb,smbphy}. The running coupling constant
for the two step interaction is known exactly \cite{jjphysa,jjb,smbphy} and
is given by \begin{equation} L\frac{\partial u_2}{\partial L} = (2-d) u_2 -
u_2^2/2\pi.\label{eq:beta2} \end{equation} The flow of $u_2$ is controlled,
for $d<2$, by the two fixed points $u_2^* = 0$ (unstable) and $u_2^* = 2\pi
(2-d)$ (stable).  The unstable FP corresponds to the transition point for
two chains and the tricritical point in the many chain case, while the
stable FP describes the repulsive steps acting like fermions in $d=1$.  For
$u_2^*=0$, the RG equation for $u_3$ is given by \cite{jjb,lassprl}
\begin{equation} L\frac{\partial u_3}{\partial L} = -c_3 u_3^2,
\end{equation} with only the fluctuation contribution in the higher order of
$u_3$ at $d=1$.\cite{comm1} Around the unstable fixed point for $u_2$, for
small deviations, the effective coupling is given by $u_2(L) \sim u_2 L$ for
$d=1$, so that the renormalized but not rescaled coupling constant is just
$v_2$.  In contrast, the renormalized three step interaction at $L\sim
\rho^{-1}$ gives $v_3(L) \sim v_3/\ln \rho$.  Substitution of these changes
the $\rho^3$ term of the free energy of Eq. \ref{eq:mffree} to $\rho^3/\ln
\rho$. The shape of the coexistence curve is then \begin{equation} v_2 \sim
\rho/\ln \rho,\quad {\rm i.e.}\quad \rho \sim \mid T - T_t\mid \ln \mid T
-T_t\mid.  \end{equation} We see that fluctuations produce a zero slope
coexistence curve, though the coexistence exponent $\beta$ is still 1, the
mean field value\cite{nagle}.  For the high temperature phase, the system is
described by the stable FP $u_2^* = 2-d$, and $v_2(L) \sim u_2^* \rho$
yielding the famous $\rho^3$ term that produces the
Kasteleyn-Pokrovsky-Talapov transition.\cite{dennijs,fisher}

We can also predict the behaviour right at the tricritical point.  With
$v_2=0$, the analogue of the Pokrovsky-Talapov transition would involve only
the three body repulsive interaction. The relevant behaviour comes from the
minimization of the free energy $F=(f_0 - \mu) \rho + v_3 \rho^3/\log
\rho$. Therefore, the step density at tricriticality behaves like $\rho \sim
\mid \mu- f_0\mid^{1/2} (\log \mid \mu - f_0\mid)^{1/2}$ with $f_0 <0$. The
exponent is the same Pokrovsky-Talapov one but with an additional log
correction (which may be hard to detect).

It is possible to have higher order multicritical points with just $v_2$ and
$v_m>0$ involving an $m$ step repulsive interaction ($m>3)$.  An exponent of
$\beta = 1/2$ can be recovered \cite{jjb} for $m=4$ in an mean field way
because $d=1$ is above the upper critical dimension of $v_m$.  Such a
multicritical point requires $v_3=0$ and with nonintersecting steps, it
seems very unlikely that this will happen.

Let us now come to the long range repulsion case, Eq. \ref{eq:hamb}.  LR
interactions are special by virtue of their singular nature.  A
renormalization group transformation is analytic in nature and, therefore,
can never generate a singular potential.  A corollary of this is that the
two body LR interaction does not get renormalized but affects the
renormalization of the short range (non singular) pair potential.  Such a
renormalization is going to change the exponent of $\rho$ in the $v_2$ term
of Eq.  \ref{eq:mffree}, and, therefore, the nature of the coexistence
curve. The RG approach for this case is also available in the literature,
and we quote the results \cite{sut,kol}.  Defining $u(L) = a [v_2(L) L^{2-d}
+ h(L) A ]$, $g(L) = 2 K_d A h(L)$, where $A=2^{d-1} \pi^{d/2} \Gamma(d/2)$,
$K_d$ is the surface area of a unit $d-$dimensional sphere, and $a$ is a
(system dependent) constant, the recursion relations from Ref \cite{kol} are
\begin{equation} L\ du/dL = - (u - u_s^*) ( u - u_u^*), \ \ {\rm and} \ \
dg/dL =0, \end{equation} where $u_{s,u}^*=\{ 2- d \pm [ (d-2)^2 +
4g]^{1/2}\}$.  Since $g$ is marginal, and the FPs for $u$ depend on the long
range coupling $g$, we find a nonuniversal behaviour.  The unstable FP
$u_u^*$ describes the tricritical point, so that linearizing around it, we
can determine the effective coupling that goes in the free energy, provided
$g \leq 3/4$.  For two chains with $g >3/4$, the binding transition is first
order \cite{kol}.  We assume that the tricritical point, as an end point of
the coexistence curve, has some critical nature. Therefore, $g\leq 3/4$.  A
straightforward analysis then gives $\Delta u(L\sim \rho^{-1}) \sim \mid T -
T_t \mid \rho^{1-x}$, where $x = u_s^* - u_u^* = [ (d-2)^2 + 4g]^{1/2}$.
This gives, via Eq. \ref{eq:mffree} a coexistence curve $\rho \sim \mid T -
T_t\mid ^{1/x}$, where, to repeat, $x$ is a nonuniversal number.  In order
to achieve consistency with experiment, one requires $x \approx 2$, which in
turn requires $g\approx 3/4$.  The RG analysis of Ref \cite{kol}, as already
mentioned, also shows $g=3/4$ is a very special point, corresponding to an
``upper critical dimension" case.  Furthermore, for $g=3/4$, with hard core
repulsion, log corrections are expected, which are not captured in the
simple RG analysis \cite{kol,lipow2}.  We conclude that if $h$ of
Eq. \ref{eq:hamb} happens to be close to $3/4\pi$, then the coexistence
exponent $\beta$ will be close to $1/2$, and if $h = 3/4\pi$, $\beta=1/2$
with additional log corrections.

In both the SR and LR cases, since the free energy is known, the surface
stiffness can also be calculated.  Following Ref. \cite{songb}, we find that
at a given $T>T_t$, for the short range case, the surface stiffness
approaches the free fermion value in a singular fashion $\sim 1/\mid \ln
\theta/d \mid $ with the misorientation while for the long range case, the
free fermion value is reached from above in a $\theta$ independent way.

To summarize, we considered two different scenarios both of which give a
zero curvature coexistence curve.  The purely short range interaction
however only gives a log correction to the mean field exponent which seems
to be far off from the experiment.  The long range case predicts a
nonuniversal value and the observed exponent seems to suggest that Si(113)
surface with a miscut towards [001] is at or close to a very special point
for the long range interaction.  It is rather striking that the very first
system that showed the tricritical point also corresponds to the very
special point for the long range interaction.  We are not sure whether it is
just an accident or a general rule.  If an accident, then other
orientational phase diagrams should be studied experimentally to verify the
claim of nonuniversality (and may be a simple verification of RG in
statistical mechanics).  If not an accident, we wonder why nature chooses to
be at the threshold.

I thank Sutapa Mukherji for many helpful discussions.  The relevance of the
long range case was suggested by M. L\"assig whom I also thank for
discussions at the initial stage of this work.  This work is partially
supported by DST SP/S2/M-17/92.

\begin{figure} 
\caption{(a) Schematic diagram
of the steps of equal height $d$.  The misorientation $\theta$ is related to
the density $\rho$, as $\tan \theta = d/l = \rho d$. (b) The steps viewed
from above. (c) Schematic phase diagram.  } 
\end{figure} 


\begin{references}
\bibitem{comm1} In case $u_2$ is not zero, there is a finite probability
that three interacting chains come together (reunion probability)
\cite{fisher,suta} and such reunions can renormalize the three
body interaction. \cite{lassprl}
\end{references}
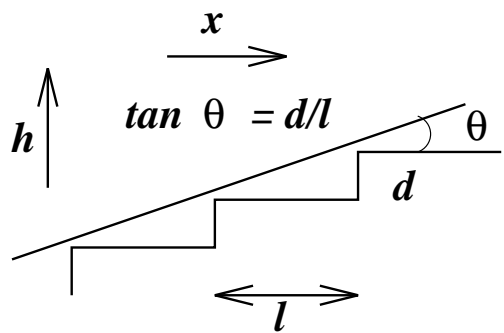
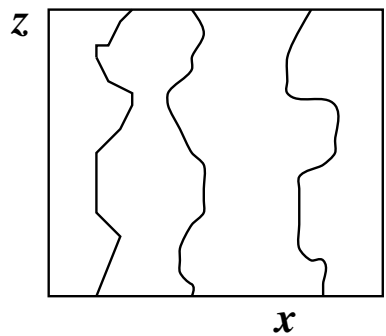
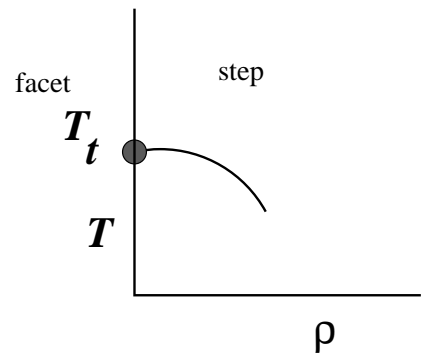
\end{document}